\documentclass[aoas,MSNbibl,nameyear,dvips]{arximspdf}
\usepackage{multirow}
\usepackage{graphicx}

\doi{10.1214/13-AOAS703} 
\volume{8}
\issue{1}
\pubyear{2014}
\firstpage{1}
\lastpage{18}
\setattribute{copyright}{owner}{In the Public Domain}

\makeatletter
\newcommand{\rrvert}{\vert}
\newcommand{\llvert}{\vert}
\newcommand{\bx}{\mathbf{x}}
\newcommand{\bB}{\mathbf{B}}
\newcommand{\bP}{\mathbf{P}}
\newcommand{\bZ}{\mathbf{Z}}

\newcommand{\tbeta}{\tilde{\beta}}
\newcommand{\hf}{\hat{f}}

\newcommand{\balpha}{\bolds{\alpha}}
\newcommand{\bbeta}{\bolds{\beta}}
\newcommand{\bsbeta}{\bolds{\beta}}

\newcommand{\E}{\mathrm{E}}

\newcommand{\logit}{\operatorname{logit}}
\makeatother

\begin{document}
\begin{frontmatter}

\title{Stochastic identification of malware with dynamic~traces\thanksref{T1}}
\runtitle{Malware detection with dynamic traces}
\thankstext{T1}{Supported in part by Los Alamos National Security,
LLC (LANS), operator of the Los Alamos National Laboratory under
Contract No.
DE-AC52-06NA25396 with the U.S. Department of Energy.
This paper is published under LA-UR 12-01317.}

\pdftitle{Stochastic identification of malware with dynamic traces}

\begin{aug}
\author[A]{\fnms{Curtis} \snm{Storlie}\thanksref{t1}\corref{}\ead[label=e1]{storlie@lanl.gov}},
\author[A]{\fnms{Blake} \snm{Anderson}\thanksref{t1}\ead[label=e2]{banderson@lanl.gov}},
\author[A]{\fnms{Scott} \snm{Vander Wiel}\thanksref{t1}\ead[label=e3]{scottv@lanl.gov}},\break
\author[D]{\fnms{Daniel} \snm{Quist}\thanksref{t2}\ead[label=e4]{daquist@Bechtel.com}}, %
\author[A]{\fnms{Curtis} \snm{Hash}\thanksref{t1}\ead[label=e5]{chash@lanl.gov}} %
\and
\author[F]{\fnms{Nathan} \snm{Brown}\thanksref{t3}\ead[label=e6]{ndbrown@nps.edu}} %
\runauthor{C. Storlie et al.}
\affiliation{Los Alamos National Laboratory\thanksmark{t1},
Bechtel Corporation\thanksmark{t2}\break
and Naval Postgraduate School\thanksmark{t3}}
\address[A]{C. Storlie\\
B. Anderson\\
S. Vander Wiel\\
C. Hash\\
Los Alamos National Laboratory\\
Los Alamos, New Mexico 87545\\
USA\\
\printead{e1}\\
\phantom{E-mail:\ }\printead*{e2}\\
\phantom{E-mail:\ }\printead*{e3}\\
\phantom{E-mail:\ }\printead*{e5}} 
\address[D]{D. Quist\\
Bechtel Corporation\\
San Francisco, California 94105\\
USA\\
\printead{e4}}
\address[F]{N. Brown\\
Naval Postgraduate School\\
Monterey, California 93943\\
USA\\
\printead{e6}}
\end{aug}

\received{\smonth{4} \syear{2013}}
\revised{\smonth{11} \syear{2013}}

%
\begin{abstract}
A novel approach to malware classification is introduced based on
analysis of instruction traces that are collected dynamically from the
program in question. The method has been implemented online in a
sandbox environment (i.e., a security mechanism for separating running
programs) at Los Alamos National Laboratory, and is intended for
eventual host-based use, provided the issue of sampling the
instructions executed by a given process without disruption to the user
can be satisfactorily addressed. The procedure represents an
instruction trace with a Markov chain structure in which the transition
matrix, $\bP$, has rows modeled as Dirichlet vectors. The malware class
(malicious or benign) is modeled using a flexible spline logistic
regression model with variable selection on the elements of $\bP$, which
are observed with error. The utility of the method is illustrated on a
sample of traces from malware and nonmalware programs, and the results
are compared to other leading detection schemes (both signature and
classification based). This article also has supplementary materials
available online.
\end{abstract}

%

\pdfkeywords{Malware detection,
classification,
elastic net,
Relaxed Lasso,
Adaptive Lasso,
logistic regression,
splines,
empirical Bayes}

\end{frontmatter}

\section{Introduction}

Malware (short for malicious software) is a term used to describe a
variety of forms of hostile, intrusive or annoying software or program
code. It was recently estimated that 30\% of computers operating in the
US are infected with some form of malware [\citet{Panda12}]. More than 286
million unique variants of malware were detected in 2010 alone [\citet
{Symantec2010}], and it is widely believed that the release rate of
malicious software is now far exceeding that of legitimate software
applications [\citet{Symantec2008}]. A large majority of the new malware
is created through simple modifications to existing malicious programs
or by using code obfuscation techniques such as a \emph{packer} [\citet
{Royal06}]. A packer compresses a program much the same way a compressor
like Pkzip does, then attaches its own decryption/loading stub which
``unpacks'' the program before resuming execution normally at the
program's original entry point (OEP).

\subsection{Review of malware detection}
Malicious software is growing at such a rate that commercial antivirus
vendors (AV) are not able to adequately keep up with new variants.
There are two methods for antivirus scanners to implement their
technology. The first is via a static signature scanning method, which
uses a sequence of known bytes in a program and tests for the existence
of this sequence. The second method of detection is to use generic or
heuristic detection technologies.
Unfortunately, even though most of the new malware is very similar to
known malware, it will often not be detected by signature-based
antivirus programs [\citet
{Christodorescu03staticanalysis,Perdisci06misleadingworm}] until the
malware signature eventually works its way into the database, which can
take weeks or even longer. Further, in a recent study [\citet{AVcomp11}],
detection of new malware threats (i.e., those not yet in the signature
database) was found to be substantially less than the ideals touted by
AV company product literature.

Because of the susceptibility to new malware, classification procedures
based on statistical and machine learning techniques have been
developed to classify new programs. These methods have generally
revolved around $n$-gram analysis of the static binary or dynamic trace
of the malicious program
[\citet
{Reddy06gramanalysis,Reddy06NewMaliciousCode,Stolfo07TowardsStealthy,Dai09Efficient}],
and some very promising results have come from a Markov chain
representation of the program trace [\citet{Anderson11GraphBasedMalware}].

The data sources used to classify programs include binary files [\citet
{Kolter06Learning,Reddy06NewMaliciousCode}], binary disassembled files
[\citet{Bilar07Opcodes,Shank2010Malware}], dynamic system call traces
[\citet
{Bayer06Dynamic,Hofmeyr98Intrusion,Rieck11Automatic}] and, most
recently, dynamic instruction traces [\citet
{Anderson11GraphBasedMalware,Dai09Efficient}]. Although some success has
been achieved by using disassembled files, this cannot always be done,
particularly if the program uses an unknown packer, and therefore, this
approach has similar shortcomings to the signature-based methods.

Here a similar path is taken to that in \citet
{Anderson11GraphBasedMalware} where they use the dynamic trace from
many samples of malware and benign programs to train a classifier. A
dynamic trace is a record of the sequence of instructions executed by
the program as it is actually running. Dynamic traces can provide much
more information about the true functionality of a program than the
static binary, since the instructions appear in exactly the order in
which they are executed during operation. The drawback to dynamic
traces is that they are difficult to collect for two reasons: (i) the
program must be run in a safe environment
and (ii) malware often has self-protection mechanisms designed to guard
against being watched by a dynamic trace collection tool, and care must
be taken to ensure the program is running as it would under normal
circumstances.

For this paper a modified version of the Ether malware analysis
framework [\citet{Dinaburg2008}] was used to perform the data collection.
Ether is a set of extensions on top of the Xen virtual machine.
Ether uses a tactic of zero modification to be able to track and
analyze a running system. Zero modification preserves
the sterility of the infected system, and limits the methods that
malware authors can use to detect if their malware is being analyzed.
Increasing the complexity
of detection makes for a much more robust analysis system.

Collecting dynamic traces can be slow due to all of the built-in
functionality of Ether to safeguard against a process altering its
behavior while being watched. Collection of traces can also be
performed with Intel's Pin tool [\citet{Luk05,Skaletsky10}], which is
faster and more stable than Ether. However, there is some concern that
Pin is more easily detected by the program being monitored. In either
case, it is an engineering hurdle to develop a software/hardware
solution that would be efficient enough to collect traces on a host
without disruption to the user. This problem is being investigated,
however, the current implementation is sufficient for application on an
enterprise network using a sandbox environment (i.e., the program is
passed along to the user that requested it, while being run on a
separate machine devoted to analysis) [\citet{Goldberg96}]. There are
several commercial sandbox tools available (e.g., FireEye, CW Sandbox,
Norman Sandbox, Malwr, Anubis, \ldots) that are used by many institutions
in a similar manner, for example.

The proposed methodology (using Pin for trace extraction) has been
inserted into this process at Los Alamos National Laboratory (LANL) and
now allows for a more robust approach to analyzing new threats as they
arrive at the network perimeter. To be clear, the extensive results and
comparisons presented in this paper used the Ether tool for trace
collection. However, Intel's Pin tool was adopted for trace extraction
to conduct model training and classification in the actual
implementation in LANL's sandbox since it was much more stable and
therefore better for use in an automated environment. While this
sandbox implementation allows the possibility of infection on an
individual user's machine, it is still a major advantage to know that a
machine has been infected with malware so that the appropriate action
can be taken. For example, it is far better to clean up a few machines
and label that file as malware instantly for other machines (i.e.,
blacklist it forever using signature-based tools) than the alternative
of not knowing about the infection until some time much later.

\subsection{Goals of this work}

The two main goals of this work are then to (i) classify malware with
high accuracy for a fixed false discovery rate (e.g., 0.001) and
(ii)~provide an online classification scheme that can determine when enough
trace data has been collected to make a decision.

As mentioned previously, this work builds upon that of \citet
{Anderson11GraphBasedMalware}, but is different in several important
ways. Most notably, goal (ii) is tackled here, but also goal (i) is
achieved in a quite different manner through categorization of
instructions and a different classification procedure, as described in
Section~\ref{secmethod}. In order to accomplish (i), we propose a
logistic regression framework using penalized splines. Estimation of
the large number of model parameters is performed with a Relaxed
Adaptive Elastic Net procedure, which is a combination of ideas from
the Relaxed Lasso [\citet{Meinshausen07}], Adaptive Lasso [\citet{Zou06b}]
and the Elastic Net [\citet{Zou05}].
To accomplish (ii), we allow the regression model to have measurement
error in the predictors in order to account for the uncertainty in a
dynamic trace with a small number of instructions.
Initial results indicate the potential for this approach to provide
some excellent and much needed protection against new malware threats
to complement the traditional signature-based approach.

The rest of the paper is laid out as follows. Section~\ref{sectracedata} describes the dynamic trace data and how it will be
used in the regression model. In Section~\ref{secmethod} the
underlying classification model and estimation methodology are
presented. The classification results on five minute dynamic traces are
presented in Section~\ref{secclassresults}. Finally, an illustration
of how the method could be applied in practice in an online
classification analysis is provided in Section~\ref{secexample}.
Section~\ref{secconclusion} concludes the paper. The supplementary
document to this paper [\citet{Storlie13malwaresupp}], available online,
also presents some preliminary work on the clustering of malware.

\section{Dynamic trace data}
\label{sectracedata}

As mentioned previously, a modified version of the Ether malware
analysis framework [\citet{Dinaburg2008}] was used to collect the dynamic
trace data. A dynamic trace is the sequence of processor instructions
called during the execution of a program. This is in contrast to a
disassembled binary \emph{static trace} which is the order of
instructions as they appear in the binary file. The dynamic trace is
generally believed to be a more robust measure of the program's
behavior since
code packers can obfuscate functionality from analysis of static traces.
Other data can be incorporated as well (e.g., presence of a packer,
system calls, file name and location, static trace, \ldots). The framework
laid out in Section~\ref{secmethod} allows for as many data sources or
features as one may wish to include.

In order to make efficient use of the dynamic trace, it is helpful to
think of the instruction sequence as a Markov chain. This
representation of the sequence has been shown to have better
explanatory power than related $n$-gram methods [\citet
{Anderson11GraphBasedMalware,Shafiq08}]. To this end, the instruction
sequence is converted into a transition matrix $\bZ$, where
\[
Z_{jk}=\mbox{number of direct transitions from instruction $j$ to
instruction $k$}.
\]
Estimated transition probabilities $\hat\bP$ are obtained from counts
$\bZ$, where
\[
P_{jk}= \Pr\{\mbox{next instruction is $k$} \mid \mbox{current
instruction is $j$}\}.
\]
The elements of $\hat\bP$ are then used as predictors to classify
malicious behavior. This entire procedure is described in more detail
in Section~\ref{secmethod}. The $Z_{jk}$ are 2-grams, while the
estimated $P_{jk}$ are essentially a scaled version of the 2-grams,
that is, the relative frequency of going from state $j$ to state $k$
given that the process is now in state $j$. These quantities ($Z_{jk}$
and $P_{jk}$) are usually substantially different in this case, since
not all states are visited with similar frequencies.
\citet{Anderson11GraphBasedMalware} used estimated $P_{jk}$ from
dynamic traces (with the state space consisting of Intel instructions
observed in the sample) as features in a support vector machine. They
found that using the $P_{jk}$ provided far better classification
results for malware than using 2-grams (or $n$-grams in general). This
is likely due to the fact that sometimes informative transitions $j
\rightarrow k$ may occur from a state $j$ that is rarely visited by a
particular program, but when it is visited, it tends to produce the $j
\rightarrow k$ transition prominently. Such situations will be measured
very differently with $P_{jk}$ versus~$Z_{jk}$.

There are hundreds of commonly used instructions in the Intel
instruction set, and thousands of distinct instructions overall. A
several thousand by several thousand matrix of transitions, resulting
in millions of predictors, would make estimation difficult. More
importantly, many instructions perform the same or similar tasks (e.g.,
``add'' and ``subtract''). Grouping such instructions together in a
reasonable way not only produces a faster method but also provides
better explanatory power versus using all distinct Intel instructions
in our experience. This is also illustrated via classification
performance in Section~\ref{secclassresults}.

Through collaboration with code writers familiar with assembly
language, we have developed four categorizations of the Intel
instructions, ranging from coarse groupings to more fine:
\begin{itemize}
\item Categorization 1 (8 classes $\rightarrow$ 64 predictors):\\
\emph{math}, \emph{logic}, \emph{priv}, \emph{branch}, \emph
{memory}, \emph{stack}, \emph{nop}, \emph{other}

\item Categorization 2 (56 classes $\rightarrow$ 3136 predictors):\\
\emph{asc}, \emph{add}, \emph{and}, \emph{priv}, \emph{bit},
\emph{call}, \emph{math\_other}, \emph{movc}, \emph{cmp}, \emph
{dcl}, $\ldots$
\item Categorization 3 (86 classes $\rightarrow$ 7396 predictors):\\
Python Library ``pydasm'' categories for Intel
instructions
\item Categorization 4 (122 classes $\rightarrow$ 14,884 predictors):\\
pydasm categories for instructions with \emph{rep
instruction-x} given its own class distinct from \emph
{instruction-x}.
\end{itemize}
%

\begin{figure}

\includegraphics{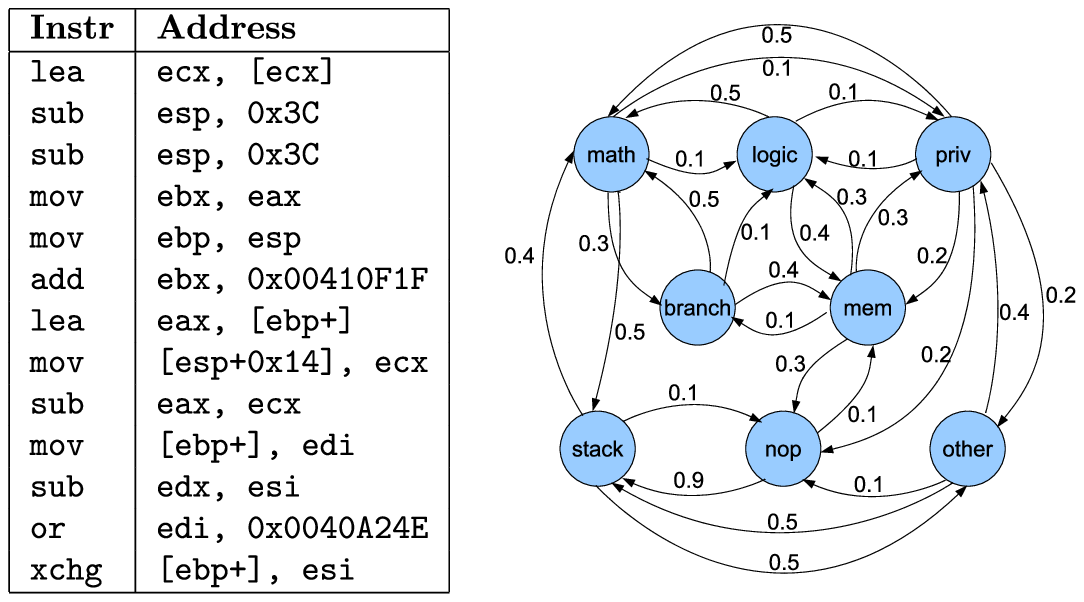}

\caption{Markov chain transition probability representation of a
dynamic instruction trace: (left) the first several lines from a
dynamic trace output (i.e., instruction and location acted on in
memory, which is not used), (right) a conceptual conversion of the
instruction sequence into categorization 1 transition probabilities.}
\label{figMCrepresentation}
\end{figure}

Figure~\ref{figMCrepresentation} displays a conceptualization of the
Markov chain transition probability representation of a dynamic
instruction trace. The graph on the right has eight nodes corresponding
to the eight categories of categorization 1, where the edges correspond
to transition probabilities from one instruction category to the next
for the given program. The location (i.e., address) that each
instruction acted on in memory is not used in this analysis, since
these locations are not consistent from one execution of the program to another.

The data set used contains dynamic traces from 18,942 malicious and
3046 benign programs, respectively, for a total of 21,988 observations.
The malicious sample was obtained via a random sample of programs from
the website \url{http://www.offensivecomputing.net/}, which is a
repository that collects malware instances in cooperation with several
institutions.
Malware samples are acquired by the Offensive Computing Website through
user contributions, capture via mwcollectors and other honeypots (i.e.,
\emph{traps} set up on a network for the purpose of collecting malware
or other information about possible network attacks), discovery on
compromised systems, and sharing with various institutions. Admittedly,
this is not a truly random sample from the population of all malicious
programs that a given network may see, but it is one of the largest
publicly available malware collections on the Internet [\citet{Quistweb}].

Obtaining a large sample of ``known to be clean'' or benign programs is
more difficult. If a program is just not deemed to be malware by AV, it
should not be given a definite ``clean bill of health,'' as will be
clear from the results of Section~\ref{secclassresults}. Hence, to
obtain a large sample of benign programs, a collection was gathered of
many programs that were running on LANL systems during 2012. If these
programs passed through a suite of 25 AV engines as ``clean,'' then
they were treated as benign for this paper. There is a far lower rate
of malware on LANL systems then in the ``wild.'' Therefore, if a
program that was on a LANL system also passes through the various AV
programs as clean, it is probably safe to deem it benign.

Each of the observations was obtained from a 5 minute run of the
program. Originally there were 19,156 malicious and 3157 benign
programs, but any program with less than 2000 instructions executed
during the five minutes was removed from the data set. The rationale
behind this was that some benign processes remain fairly idle, waiting
for user interaction. Since such programs produce very short traces and
are not representative of the kind of programs that require scanning,
it was decided to remove them from the training set. The data set used
in this analysis (dynamic trace files obtained via Ether) is available
upon request.






\section{A statistical model for malware classification}
\label{secmethod}

Elements of an estimated probability transition matrix for the dynamic
trace of each program $\hat\bP$ are used as predictors to classify a
program as malicious or benign. Two main goals for the classification
model are to (i) screen out most predictors to improve performance and
allow certain transitions to demonstrate their importance to
classification, and (ii) explicitly account for uncertainty in ${\hat
\bP}$ for online classification purposes because this uncertainty will
have a large impact on the decision until a sufficiently long trace is obtained.

\subsection{Logistic spline regression model}

For a given categorization (from the four categorizations given in
Section~\ref{sectracedata}) with $c$ instruction categories, let
$\bZ
_i$ be the transition counts between instruction categories for the
$i$th observation (an observation $\bZ_i$ is made once on each
program). Let $B_i$ be the indicator of maliciousness, that is, $B_i =
1$ if the $i$th observation is malicious and $B_i = 0$ otherwise. For
the initial model fit discussion in this section, we take $\bP_i$ to be
fixed at an estimated value~$\hat\bP_i$. The training set has
observations where the traces are long enough so that there is very
little variability in $\hat\bP_i$.
In the results of Section~\ref{secclassresults} we specifically take
$\hat\bP_i$ to be the posterior mean [i.e., $\E(\bP_i \mid\bZ_i)$],
assuming a symmetric $\operatorname{Dirichlet}(\nu)$ prior for each row of $\bP_i$
(i.e., an independent Dirichlet distribution with parameter vector
$[\nu
,\nu,\ldots,\nu]$ was assumed for each row of $\bP_i$).
In the analysis presented in this paper $\nu=0.1$ was used. However, in
Section~\ref{secexample} the assumption of a fixed known $\bP_i$
(equal to $\hat\bP_i$) is relaxed and a simple approach is described to
account for the uncertainty inherent in $\bP_i$ when making decisions
based on shorter traces.

The actual predictors used to model the $B_i$ are
%
\begin{equation}
\bx_i= \bigl[\logit(\hat{P}_{i,1,1}), \logit(\hat
{P}_{i,1,2}),\ldots ,\logit(\hat{P}_{i,c,c-1}), \logit(
\hat{P}_{i,c,c}) \bigr]' \label{eqpredictors}
\end{equation}
for $i=1,\ldots,n$, where $\hat{P}_{i,j,k}$ is the $(j,k)$th entry of
the $\hat\bP_i$ matrix, and each component of the $\bx_i$ is scaled to
have sample mean 0 and sample variance 1, across $i=1,\ldots,n$. The
scaling of the predictors to a comparable range is a standard practice
for penalized regression methods [\citet{Tibs96,Zou05}]. We then use the model
%
\begin{equation}
\operatorname{logit}\bigl[\Pr(B=1)\bigr] = f_{\bsbeta}(\bx) =
\beta_0 + \sum_{s=1}^{c^2} \sum
_{l=1}^{K+1} \beta_{s,l}
\phi_{s,l}(x_s), \label{eqlogisticmodel}
\end{equation}
where the basis functions $\phi_{s,1},\ldots, \phi_{s,K+1}$ form a
linear spline with $K$ knots at equally spaced quantiles of $x_s$,
$s=1,\ldots,c^2$ (and $c^2$ is the number of elements in the $\hat{\bP}$
matrix). That is, $\phi_{s,1}(x_s) = x_s$ and
\[
\phi_{s,l}(x_s) = |x_s - \xi_{s,l}
|_+ \qquad\mbox{for } l=2,\ldots,K+1,
\]
where $\xi_{s,l}$ is the $(l-1)$th knot for the $s$th predictor, and
$|x|_+ = x$ if $x>0$ and 0 otherwise.

Pairwise products of the $\phi_{s,l}(x)$ can also be included to create
a \emph{two-way interaction} spline for $f(\bx)$. A compromise which is
more flexible than the additive model in (\ref{eqlogisticmodel}), but
not as cumbersome as the full two-way interaction spline, is to use a
model which includes multiplicative interaction terms, that is,
%
\begin{equation}
f_{\bsbeta}(\bx) = \beta_0 + \sum
_{s=1}^{c^2} \sum_{l=1}^{K+1}
\beta_{s,s,l} \phi_{s,s,k}(x_s) + \sum
_{s=1}^{c^2-1} \sum_{t=s+1}^{c^2}
\sum_{l=1}^{K+1} \beta_{s,t,l}
\phi_{s,t,l}(x_s x_t), 
\label{eqlogisticmodel2}
\end{equation}
where $\phi_{s,t,1},\ldots, \phi_{s,t,K+1}$ form a linear spline with
$K$ knots at equally spaced quantiles of $x_s x_t$ for $s \neq t$ (and
at equally spaced quantiles of $x_s$ for $s=t$).
The model in (\ref{eqlogisticmodel2}) with $K=5$ is ultimately the
route taken for implementation of the detection scheme on our network.
This model has potentially a very large number of parameters in this
application [$\sim$30 million $\beta $'s for the interaction model in
(\ref{eqlogisticmodel2}) when using categorization 2]. Thus, an
efficient sparse estimation procedure is necessary and is described
next in Section~\ref{secrelaxedEN}.

\subsection{Relaxed Adaptive Elastic Net estimation}
\label{secrelaxedEN}

In order to estimate the large number of parameters in (\ref
{eqlogisticmodel2}), a combination of the Elastic Net [\citet{Zou05}],
Relaxed Lasso [\citet{Meinshausen07}] and Adaptive Lasso [\citet{Zou06b}]
procedures was used. The Elastic Net is efficient and useful for
extremely high-dimensional predictor problems $p\gg n$. This is in part
because it can ignore many unimportant predictors (i.e., it sets many
of the $\beta_{s,t,l}\equiv0$). The Elastic Net, Relaxed Lasso and
Adaptive Lasso procedures are reviewed below, then generalized for use
in this application.

The data likelihood is
%
\[
L(\bbeta; \bB) = \prod_{i=1}^n \bigl[
\logit^{-1} \bigl(f_{\bsbeta
}(\bx_i) \bigr)
\bigr]^{I_{B_i=1}} \bigl[1-\logit^{-1} \bigl(f_{\bsbeta}(
\bx_i) \bigr) \bigr]^
{I_{B_i=0}},
\]
where $\bB= [B_1,B_2,\ldots,B_n]^t$.
The Elastic Net estimator is a combination of ridge regression and
Lasso [\citet{Tibs96}], that is, it seeks the $\bbeta$ that minimizes
%
\begin{equation}
\log L(\bbeta; \bB) + \lambda \Biggl\{ \rho\sum_{s=1}^{c^2}
\sum_{t=s}^{c^2} \sum
_{l=1}^{K+1} \beta_{s,t,l}^2 + (1-
\rho) \sum_{s=1}^{c^2} \sum
_{t=s}^{c^2} \sum_{l=1}^{K+1}
\llvert \beta _{s,t,l} \rrvert \Biggr\} \label{eqEN}
\end{equation}
for given tuning parameters $\lambda>0$ and $\rho\in[0,1]$ which are
typically chosen via $m$-fold cross-validation (CV).
For the linear spline model of (\ref{eqlogisticmodel2}), the penalty
on $\beta_{s,t,l}^{ 2}$ and $\llvert \beta_{s,t,l} \rrvert$ corresponds
to a penalty on the overall trend and the change in slope at the knots
(i.e., encourages ``smoothness'').
Another benefit to the Elastic Net is that fits for many values of
$\lambda$ are obtained at the computational cost of a single least
squares fit [i.e., $O(p^2)$] via the Least Angle Regression (LARS)
algorithm [\citet{Efron04b}].

The Relaxed Lasso and Adaptive Lasso both emerged as procedures
developed to counteract the over-shrinking that occurs to the nonzero
coefficients when using the Lasso procedure in high dimensions.
The Relaxed Lasso can be thought of as a two-stage procedure, where the
Lasso procedure (i.e., the Elastic Net estimator with $\rho=0$) is
applied with $\lambda=\lambda_1$, then the Lasso is applied again to
only the nonzero coefficients with $\lambda=\lambda_2$, where typically
$\lambda_1>\lambda_2$.

The Adaptive Lasso is also a two-stage procedure where an initial
estimate of $\beta_{s,t,l}$ is obtained via unregularized maximum
likelihood estimates or via ridge regression (if $p>n$). In the second
step, the Lasso is applied with a penalty that has each term weighted
by the reciprocal of the initial estimates, $\tbeta_{s,t,l}$.

This motivates the following three-step approach taken to estimate the
coefficients of the logistic spline model in (\ref{eqlogisticmodel2}):\vspace*{6pt}

\textit{Algorithm} 1: Estimation procedure.
\begin{longlist}[\emph{Step} 1:]

\item[\emph{Step} 1:] Screen the predictors $x_s$ for
importance (i.e., conduct variable selection) using a linear logistic model
\[
\operatorname{logit} \bigl[\operatorname{Pr}(B = 1) \bigr] = f_1(
\bx) = \alpha_0 + \sum_{s}
\alpha_{s} x_s,
\]
with $\balpha$ estimated via Elastic Net in (\ref{eqEN}) with
$\lambda
= \lambda_1$ and $\rho=0.5$.
\item[\emph{Step} 2:] Use active predictors (i.e., those
$x_s$ with $\alpha_s \neq0$) to fit the interaction spline model of
(\ref{eqlogisticmodel2}) via the Elastic Net with $\lambda=
\lambda
_2$ and $\rho=0.5$. Denote the estimated coefficients from step 2 as
$\tbeta_{s,t,l}$.
\item[\emph{Step} 3:] Fit the interaction spline model of
(\ref{eqlogisticmodel2}) via the Adaptive Elastic Net with $\lambda
= \lambda_3$ and $\rho=\rho_3$. That is, $\hat{\bbeta}$ is given
by the
minimizer of
\end{longlist}
%
%
\begin{equation}\qquad
\log L(\bbeta; \bB) + \lambda_3 \Biggl\{ \rho_3 \sum
_{s=1}^{c^2} \sum_{t=s}^{c^2}
\sum_{l=1}^{K+1} \biggl(\frac{\beta_{s,t,l}}{\tbeta_{s,t,l}}
\biggr)^2 + (1-\rho_3) \sum_{s=1}^{c^2}
\sum_{t=s}^{c^2} \sum
_{l=1}^{K+1} \biggl\llvert \frac{\beta_{s,t,l}}{\tbeta_{s,t,l}} \biggr
\rrvert \Biggr\}. \label{eqadaptiveEN}
\end{equation}

The tuning parameters $\lambda_1$, $\lambda_2$, $\lambda_3$ and
$\rho
_3$ need to be chosen via CV. However, these parameters are tuned
individually within their respective steps of the fitting algorithm
(i.e., tune $\lambda_1$ in step 1, then tune $\lambda_2$ in step 2,
then tune $\lambda_3$ and $\rho_3$ in step~3). That is, they are not
tuned to find a global optimum in the four-dimensional space. As
mentioned above, tuning of $\lambda$ comes at no additional cost in the
LARS algorithm. So the only loop needed is in step 3 of the algorithm
to tune $\rho_3$. Additionally, the implementation of the LARS
algorithm via the R package \texttt{glmnet} allows a specification of the
maximum number of nonzero coefficients to ever be included into the
model. This max was set to 20,000 in each step of Algorithm 1 in our
implementation.

On the surface it may seem excessive to combine these three concepts,
but the extremely high dimensionality of the model in (\ref
{eqlogisticmodel2}) demands this aggressive approach. There are over
9 million parameters if using categorization 2, over 200 million
predictors if using categorization 4. The initial out-of-sample
classification results using just one of these procedures alone were
far inferior to those obtained with combined approach. For example,
overall 10-fold CV classification rates of $\sim$94\% were achieved
with the Elastic Net, Adaptive Lasso and Relaxed Lasso, respectively,
when used alone to fit the model in (\ref{eqlogisticmodel2}),
whereas overall 10-fold CV accuracies achieved using the combined
method above are $\sim$98\%, as shown in Section~\ref{secclassresults}. One could also use another sparse estimation
routine in place of Elastic Net in Algorithm 1. In fact, the results of
using the Maximum a posteriori (MAP) estimate discussed in \citet
{Taddy13} are also provided in Section~\ref{secclassresults}.

\subsection{Prior correction for sample bias}

Prior correction [\citet{Manski77,Prentice79}] involves computing the
usual logistic regression fit and correcting the estimates based on
prior information about the proportion of malware in the population of
interest $\pi_1$ and the observed proportion of malware in the sample
(or sampling probability), $\bar B$. Knowledge of $\pi_1$ can come from
some prior knowledge, such as expert elicitation or previous data.
\citet{King01} point out that, provided the estimates of the
regression coefficients [i.e., $\beta_{s,t,l}$, $j<k, l=1,\ldots,M$ in
(\ref{eqadaptiveEN})] are consistent, the following corrected
estimate is consistent for $\beta_0$:
%
\begin{equation}
\tilde{\beta}_0 = \hat{\beta}_0 - \log\biggl(
\frac{1-\pi_1}{\pi_1} \frac{\bar{B}}{1-\bar{B}} \biggr). \label{eqpriorcorrect}
\end{equation}

Prior correction will have no effect on the classification accuracy
results discussed in Section~\ref{secclassresults}, since it is just
a monotonic transformation, so there will be an equivalent threshold to
produce the same classification either way. However, it can be useful
in practice to have the estimated probability of maliciousness for a
given program provide a measure of belief of the maliciousness of the
program on a scale that incorporates the prior probability that the
code is malicious. That is, if $\pi_1$ can somehow be specified for the
given network on which the program will be executed, then prior
correction in (\ref{eqpriorcorrect}) can be useful.

\section{Classification results}
\label{secclassresults}

Let $\widehat{\Pr}(B=1 \mid\bx)$ be given by (\ref
{eqlogisticmodel2}) with $\beta_{s,t,l}$ replaced by their
respective estimates $\hat\beta_{s,t,l}$. The $i$th observation is
classified as malicious if $\widehat{\Pr}(B=1 \mid\bx_i) > \tau$ for
some threshold $\tau$ which can be selected to produce an acceptable
false discovery rate (FDR).

\subsection{Out of sample accuracy}

The classification accuracy of the proposed method is first examined on
the various categorizations. The 10-fold CV overall accuracy results
for the four different categorizations are provided in Table~\ref{tabcataccuracy}. The overall accuracy is defined as the number of
correct classifications divided by the number of programs.
Categorizations 2, 3 and 4 are generally not much different from each
other, but they all perform far better than categorization 1. In the
remainder of the results the categorization 2 is used, since it
provides the most parsimonious model among the best performing
covariate sets.

%
\begin{table}
\caption{Overall out-of-sample accuracy calculated via 10-fold CV by
category using logistic interaction spline regression with Relaxed
Adaptive Elastic Net estimation. The standard error of the respective
accuracy estimates are provided in parentheses}
\label{tabcataccuracy}
\begin{tabular*}{\textwidth}{@{\extracolsep{\fill}}lccc@{}}
\hline
\textbf{Cat 1} & \textbf{Cat 2} & \textbf{Cat 3} & \textbf{Cat 4}\\
\hline
0.923 (0.009) & 0.976 (0.004) & 0.971 (0.003) & 0.971 (0.003)\\
\hline
\end{tabular*}
\end{table}


The logistic spline regression with Relaxed Adaptive Elastic Net
estimation \emph{EN Inter Spline} is compared to various other methods
(with all methods using categorization 2 unless explicitly stated
otherwise) in Table~\ref{tabcompareacc}. The data set was partitioned
into the same 10 sets to produce the 10-fold CV results for all
methods. The entire estimation routine (including scaling the
covariates and parameter tuning) was conducted on each of the ten
training sets, then predictions on the respective test sets were obtained.

The competing methods are (i) \emph{MAP Inter Spline}---using the MAP
estimate of \citet{Taddy13} from the R-package \texttt{textir} in
place of the elastic net throughout Algorithm 1, (ii) \emph{EN
Linear$+$Int}---using a linear model with interaction instead of the
interaction spline in Algorithm 1, (iii) \emph{EN Add Spline}---using an
additive spline instead of the interaction spline in Algorithm 1, (iv)
\emph{EN Linear}---linear logistic regression estimated with Elastic Net
(i.e., step 1 of Algorithm 1), (v)~\emph{MDA}---the mixture discriminant
analysis (MDA) routine of \citet{Hastie96} (using the R package
\texttt{mda}) using two components on the set of covariates with nonzero
coefficients from the linear logistic regression elastic net, (vi)~\emph{SVM}---a support vector machine using a Gaussian kernel provided
by the
R~package \texttt{kernlab}, (vii) \emph{SVM (No Cat)}---a support vector
machine using a Gaussian kernel and using a distinct category for each
unique instruction [i.e., the approach of \citet
{Anderson11GraphBasedMalware} implemented with the author's C code],
and (viii) \emph{Antivirus 1--7}---seven leading signature-based
antivirus programs with all of their most recent updates. The predictor
screening used in conjunction with the MDA method is essential in this
case in order to avoid the numerical issues with the procedure that
occurred when using all predictors. The number of mixture components
(two) was chosen to produce the best 10-fold CV accuracy.

%
\begin{table}
\tabcolsep=0pt
\caption{Comparison of classification results using various methods.
All methods use categorization 2 with the exception of SVM (No Cat),
which assumes each unique instruction is its own unique category. All
methods had results calculated via 10-fold CV (same 10 folds were used
for each method)}
\label{tabcompareacc}
\begin{tabular*}{\textwidth}{@{\extracolsep{\fill}}lcccccc@{}}
\hline
&  &
\multicolumn{3}{c}{\textbf{Malware detection accuracy\scriptsize{\tabnoteref{t1}}}}
&\multicolumn{2}{c@{}}
{\multirow{2}{50pt}[8pt]{\centering\textbf{Compute time (min)}}} \\[-6pt]
&  & \multicolumn{3}{c}{\hrulefill}
&\multicolumn{2}{c@{}}{\hrulefill} \\
\multicolumn{1}{c}{\textbf{Detection method}} &
\multicolumn{1}{c}{\multirow{2}{40pt}[10pt]{\centering \textbf{Overall accuracy}}}
 & \textbf{1\% FDR\scriptsize{\tabnoteref{t2}}} &
\textbf{0.1\% FDR\scriptsize{\tabnoteref{t3}}} &
\multicolumn{1}{c}{$\bolds{\sim}$\textbf{0\%} \textbf{FDR\scriptsize{\tabnoteref{t4}}}} &
\multicolumn{1}{c}{\textbf{Train\scriptsize{\tabnoteref{t5}}}} &
\multicolumn{1}{c@{}}{\textbf{Predict\scriptsize{\tabnoteref{t6}}}}\\
\hline
EN Inter Spline & 0.976 & 0.817 & 0.692 & & 934 & 0.291 \\
MAP Inter Spline & 0.969 & 0.807 & 0.594 & & 265 & 0.282 \\
SVM & 0.963 & 0.597 & 0.333 & & 231 & 0.328 \\
EN Linear$+$Int & 0.961 & 0.679 & 0.373 & & \phantom{0}90 & 0.013 \\
EN Add Spline & 0.952 & 0.646 & 0.342 & & 483 & 0.187 \\
EN Linear & 0.944 & 0.564 & 0.272 & & \phantom{0}32 & 0.002 \\
SVM (No Cat) & 0.936 & 0.420 & 0.315 & & 447 & 0.929 \\
MDA & 0.926 & 0.252 & 0.115 & & \phantom{0}33 & 0.001 \\
Antivirus 1 & 0.838 &&& 0.812 & & \\
Antivirus 2 & 0.831 &&& 0.804 & & \\
Antivirus 3 & 0.825 &&& 0.797 & & \\
Antivirus 4 & 0.805 &&& 0.774 & & \\
Antivirus 5 & 0.790 &&& 0.756 & & \\
Antivirus 6 & 0.619 &&& 0.558 & & \\
Antivirus 7 & 0.287 &&& 0.172 & & \\
\hline
\end{tabular*}
\tabnotetext[1]{t1}{The number of correct malware classifications divided by
the number of malware observations.}
\tabnotetext[2]{t2}{30 out of 3046 benign programs incorrectly considered
malicious.}
\tabnotetext[3]{t3}{Three out of 3046 benign programs incorrectly considered
malicious.}
\tabnotetext[4]{t4}{There are \emph{some} false positives from signature-based
detection methods due to fuzzy matching heuristics (Antivirus 1 and
Antivirus 2 had two and one false detections, respectively, in this
data set, e.g.), but the exact FDR for these signature-based
methods is unknown.}
\tabnotetext[5]{t5}{Time in minutes to conduct the estimation of the model
including parameter tuning (via 10-fold CV) for one of the ten training
sets (i.e., using $\sim$20,000 observations) once the traces have been
collected and parsed.}
\tabnotetext[6]{t6}{Time in minutes to conduct the classification of all of
the programs in one of the ten test sets (i.e., $\sim$2200 programs)
once the traces have been collected and parsed.}
\end{table}

The names of the leading antivirus programs are not provided due to
legal concerns. It is important to recognize that these AV software
packages are using signatures (i.e., blacklists) and whitelists as well
as heuristics to determine if a program is malicious. The other (i.e.,
classification-based) methods in the table are \emph{not} using
signatures, hence, a direct comparison to the AV results is not
possible. In particular, the results of the classification approaches
would improve substantially if signatures were to be used to ensure
correct classification for some of the programs in the data set. It is
important to understand that the goal of the classification-based
methods is not to replace signature-based detection, but rather to
complement signature-based methods by providing protection from new
threats. Even without the use of signatures, the interaction spline
logistic method is competitive with signature-based methods and would
be a promising addition to AV software. The \emph{EN Linear$+$Int} method
is also competitive in terms of accuracy and at a small fraction of the
computational cost for training and prediction.

Another competing method that would ideally be used in this comparison
is the posterior mean of the probability of maliciousness (as opposed
to the MAP), as discussed in \citet{Gramacy12b} and provided by
the R package \texttt{reglogit}. In fact, when assuming a Bayesian
logistic regression model, thresholding on the posterior mean would be
Bayes optimal. In this particular problem, though, the covariate space
was too large for this approach to be practical. However, the MAP and
posterior mean will be nearly identical for a large enough sample. In
this paper, for example, we have a sample of over 20,000 and, in
practice, this number is growing everyday. In an effort to compare MAP
to posterior mean for a large sample, we ran with both approaches on
the logistic regression model with linear terms using categorization 1.
This test case was chosen because it posed no computational issues. The
MAP and posterior mean provided nearly identical results, for example,
out-of-sample accuracies were 0.911 and 0.908, respectively. This
provides some assurance that the MAP will perform similarly to the
posterior mean here, however, it is hard to know for certain how large
of a sample is needed for this difference to become negligible for the
interaction splines model on categorization 2.

\begin{figure}

\includegraphics{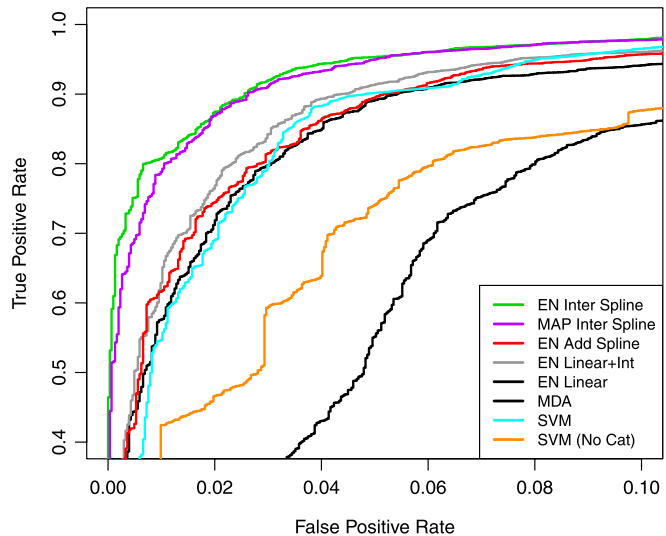}

\caption{ROC curves for the methods in Table~\protect\ref{tabcompareacc}.}
\label{figROCcurves}
\end{figure}

Figure~\ref{figROCcurves} displays the ROC curves for the various
competing methods in Table~\ref{tabcompareacc}. The antivirus
programs are excluded since there is no thresholding parameter with
which to vary the false positive rate. It is clear from Table~\ref{tabcompareacc} and Figure~\ref{figROCcurves} that the interaction
spline logistic model with Relaxed Adaptive Elastic Net estimation or
Relaxed Adaptive MAP estimation are superior to the other methods for
this classification problem. In particular, the Interaction Spline EN
procedure has an estimated out-of-sample overall error rate of only
0.024 (accuracy of 97.6\%) and still maintains a high degree of malware
identification accuracy (69.2\%) when held to a very small (0.1\%)
false positive rate. When implementing this procedure in practice, it
would be wise to actually tune the estimation procedure not to
necessarily obtain the best \emph{overall} accuracy as was done here,
but rather to specifically obtain high accuracy at identifying malware
for a small false positive rate.

\section{Online detection}
\label{secexample}

The predictors used in the logit spline model of Section~\ref{secmethod} are the elements of the probability transition matrix $\bP
$, which can only be observed (i.e., estimated) with error. This \emph
{measurement} error can be substantial for a dynamic trace with only a
small number of instructions. For online classification, it is
essential to account for this measurement error and its effect on the
probability of maliciousness. Of particular importance is determining
how long of an instruction trace is needed before a decision can be made.

\begin{figure}[b]
\centering
\begin{tabular}{@{}cc@{}}
\includegraphics{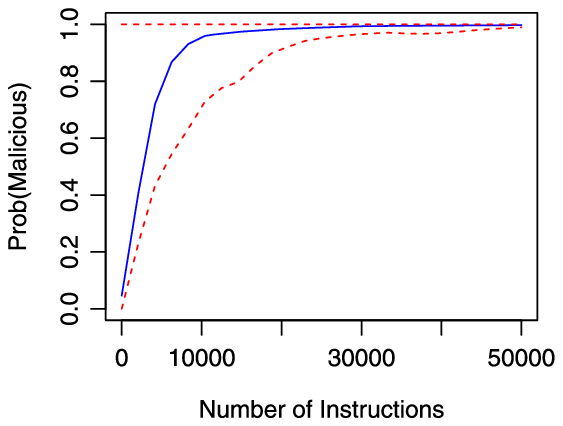}
&
\includegraphics{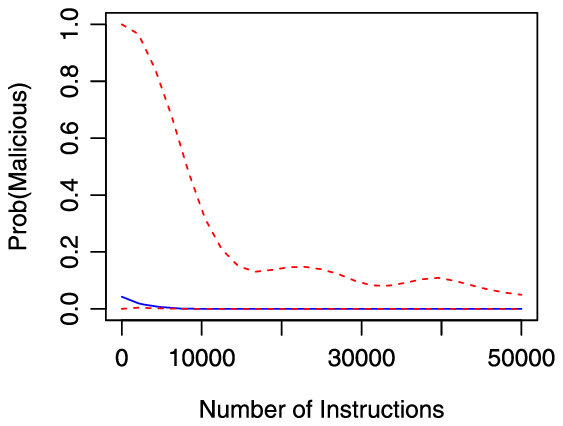}\\
\footnotesize{(a) Dynamic trace from a malicious program} &
\footnotesize{(b) Dynamic trace from a benign program}
\end{tabular}
\caption{Posterior mean of the probability of malware given the
instruction sequence for a malicious sample \textup{(a)} and benign sample \textup{(b)},
respectively, as a function of number of instructions (95\% CI
reflecting uncertainty in red).}\label{figonlineprob} 
\end{figure}

To tackle this issue, each row of $\bP$ is further modeled as a
symmetric $\operatorname{Dirichlet}(\nu)$ vector {a priori}, with each row assumed
independent. The Dirichlet distribution is a conjugate prior for $\bP$
in a Markov chain model. With this assumption, and a trace $T_{1:m}$
with $m$ instructions observed thus far, the probability of being
Malicious, $\Pr(B=1) = \logit^{-1}(\hat{f}(\bP))$, has inherent
variability (due to the uncertainty in $\bP$) that decreases as $m$
increases (i.e., as a longer instruction trace is obtained). If a given
process produces a trace $T_{1:m}$, the distribution of $\Pr(B=1)$ can
be simulated by generating draws from the posterior of $\bP$ to produce
uncertainty bands and a \emph{posterior} mean estimate $\E[\Pr(B=1)
\mid T_{1:m}]$.

This can be thought of as an empirical Bayes approach, as $f$ is
replaced with an estimate $\hf$, while only the uncertainty in $\bP$ is
recognized.
This is a good compromise, since there is a large sample available to
construct the classification model, and the uncertainty in $\Pr(B=1)$
is dominated by uncertainty in $\bP$ early on in the trace. Ideally, a
fully Bayesian version of this procedure could be implemented,
accounting for the ``measurement'' error even in the logistic
regression estimation process. However, this was attempted and proved
to be computationally infeasible for this problem. Figure~\ref{figonlineprob} demonstrates the empirical Bayes approach on the
first malicious and benign programs in the sample, respectively, using
a prior correction of $\pi_1=0.01$. There is a lot of uncertainty in
either case initially, until about 10,000 instructions are collected
(typically this takes a few seconds of runtime or less). By about
30,000 instructions the $\Pr(B=1)$ for the malicious and benign
processes are tightly distributed near one and zero, respectively. A
possible implementation for online decision making could be to classify
as malicious (or benign) according to $\Pr(B=1)>\tau$ (for some
threshold $\tau$ that admits a tolerable number of alarms per day) once
the 95\% credible interval (CI) is narrow enough (e.g., $<$0.1).

%
%
%

\begin{figure}

\includegraphics{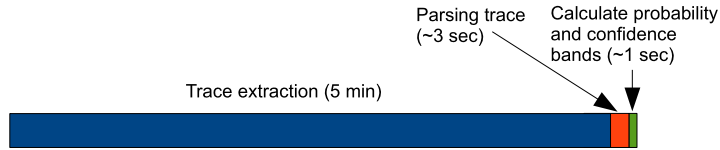}

\caption{Computation time breakdown for analysis of a particular new
program that generated $\sim\!3.5 \times10^6$ instructions in a 5 minute
trace. CIs calculated from a sample of 1000 posterior draws of $\bP$.}
\label{figtimebreakdown}
\end{figure}

Figure~\ref{figtimebreakdown} provides a breakdown of the
computational time needed for each piece of the analysis of a new
program. It is clear that most of the time is spent on the extraction
of the trace itself. This five minutes, however, can be significantly
shortened in many cases using the CI approach discussed above. With
this current computational burden, this approach is currently only
suitable for use on a network \emph{sandbox} (i.e., running on a server)
as it is passed along to the host machine. A trace extraction in-line
on the host could be feasible via a different software or hardware
solution, however, and is currently being investigated further.

Once a file is suspected of being malicious, it is often necessary to
reverse engineer (RE) the program to determine its functionality and
origin in order to know how to respond and/or how to better prevent
future infections into the computer network. The reverse engineering
process is fairly sophisticated, requiring many hours of effort from a
highly trained individual. Some preliminary work has been done to
cluster malware instances into homogeneous groups to speed up the RE
process. This work is presented in a supplemental document to this
paper [\citet{Storlie13malwaresupp}] which is available online. In that
document a novel clustering method based on a probabilistic change
similarity measure is described. When a new malware instance is
detected, it can be clustered into a homogeneous group, where perhaps
some of the group members have already been reverse engineered by an
analyst. The analyst can then use these previous efforts to more
quickly understand the nature and functionality, origin, etc.~of the
newly suspected malicious program. An example of how this would work on
a particular malware instance is also provided.








\section{Conclusions and further work}
\label{secconclusion}

The Relaxed Adaptive Elastic Net is a good framework for adding
flexibility with splines for classification in high-dimensional settings.
It avoids over-fitting to obtain superior accuracy relative to other
popular approaches for classifying new programs.
It is also possible to use a model-based classification approach that
treats the dynamic trace as a Markov chain directly, and assumes a
mixture of Dirichlet distributions for the rows of the transition
matrix $\bP$. This framework would also cluster malware samples as it
detects them. Also, \citet{Anderson12} use other features (e.g.,
static trace, file entropy, system calls) in addition to dynamic traces
(obtained via Pin) to perform classification. The approach discussed
here would easily allow for these additional features along with the
dynamic traces. Since categorization of instructions is useful, it
seems prudent to further investigate what the best possible
categorization might be.

The current framework allows for online application in a sandbox at the
perimeter of a network. Implementation of the classification procedure
in this manner is currently up and running on LANL's network and has
identified several instances of malware that were not found by any of
LANL's commercial AV tools. Thus, this approach has proven to be a
useful supplement to traditional signature-based AV. The classification
procedure runs very quickly on a given trace (once the model is
estimated, which is done offline). The largest obstacle to producing a
host-based software (i.e., running the classification procedure on an
actual user's machine as it runs the applications) is the collection of
dynamic traces in real time efficiently without disruption to the user.
The feasibility of such a collection procedure is currently being investigated.
Finally, it could be useful to incorporate change point detection in
order to allow for instances where a benign process is ``hijacked'' by
a malicious program [\citet{Cova10}].

\section*{Acknowledgments} The authors thank Joshua Neil and Mike Fisk
for their helpful discussions related to this paper, and thank two
referees and the Editor for their thoughtful comments and suggestions.

\begin{supplement}[id=suppA]
\stitle{Supplement to ``Stochastic identification and clustering of
malware with dynamic traces''}
\slink[doi]{10.1214/13-AOAS703SUPP} 
\sdatatype{.pdf}
\sfilename{aoas703\_supp.pdf}
\sdescription{This article also has a supplemental document \citet
{Storlie13malwaresupp} available online which presents preliminary work
on the clustering of malware, to aid in reverse engineering. Some
computational complexity considerations for the proposed method are
also discussed.}
\end{supplement}

%

%


\printaddresses


\begin{thebibliography}{38}


\bibitem[\protect\citeauthoryear{Anderson et~al.}{2011}]{Anderson11GraphBasedMalware}
\begin{barticle}[auto:STB|2014/01/06|10:16:28]
\bauthor{\bsnm{Anderson},~\bfnm{B.}\binits{B.}},
\bauthor{\bsnm{Quist},~\bfnm{D.}\binits{D.}},
\bauthor{\bsnm{Neil},~\bfnm{J.}\binits{J.}},
\bauthor{\bsnm{Storlie},~\bfnm{C.}\binits{C.}} \AND
\bauthor{\bsnm{Lane},~\bfnm{T.}\binits{T.}}
(\byear{2011}).
\btitle{Graph-based malware detection using dynamic analysis}.
\bjournal{Journal in Computer Virology}
\bvolume{7}
\bpages{247--258}.
\end{barticle}
\bptok{imsref}%
\endbibitem

\bibitem[\protect\citeauthoryear{Anderson et~al.}{2012}]{Anderson12}
\begin{bmisc}[auto:STB|2014/01/06|10:16:28]
\bauthor{\bsnm{Anderson},~\bfnm{B.}\binits{B.}},
\bauthor{\bsnm{Quist},~\bfnm{D.}\binits{D.}},
\bauthor{\bsnm{Brown},~\bfnm{N.}\binits{N.}},
\bauthor{\bsnm{Storlie},~\bfnm{C.}\binits{C.}} \AND
\bauthor{\bsnm{Lane},~\bfnm{T.}\binits{T.}}
(\byear{2012}).
\bhowpublished{Improving malware classification: Bridging the static/dynamic gap.
In \textit{Proceedings of the 5th ACM Workshop on Security and Artificial Intelligence}
3--14. ACM, New York.}
\end{bmisc}
\bptok{imsref}%
\endbibitem

\bibitem[\protect\citeauthoryear{Antivirus Comparatives}{2011}]{AVcomp11}
\begin{bmisc}[auto:STB|2014/01/06|10:16:28]
\borganization{Antivirus Comparatives}
(\byear{2011}).
\bhowpublished{Retrospective test (static detection of new/unknown malicious software). Available at
\texttt{\href{http://www.av-comparatives.org/images/stories/test/ondret/avc\_retro\_nov2011.pdf}{http://www.av-comparatives.org/}
\href{http://www.av-comparatives.org/images/stories/test/ondret/avc\_retro\_nov2011.pdf}{images/stories/test/ondret/avc\_retro\_nov2011.pdf}}.}
\end{bmisc}
\bptok{imsref}%
\endbibitem

\bibitem[\protect\citeauthoryear{Bayer et~al.}{2006}]{Bayer06Dynamic}
\begin{barticle}[auto:STB|2014/01/06|10:16:28]
\bauthor{\bsnm{Bayer},~\bfnm{U.}\binits{U.}},
\bauthor{\bsnm{Moser},~\bfnm{A.}\binits{A.}},
\bauthor{\bsnm{Kruegel},~\bfnm{C.}\binits{C.}} \AND
\bauthor{\bsnm{Kirda},~\bfnm{E.}\binits{E.}}
(\byear{2006}).
\btitle{Dynamic analysis of malicious code}.
\bjournal{Journal in Computer Virology}
\bvolume{2}
\bpages{67--77}.
\end{barticle}
\bptok{imsref}%
\endbibitem

\bibitem[\protect\citeauthoryear{Bilar}{2007}]{Bilar07Opcodes}
\begin{barticle}[auto:STB|2014/01/06|10:16:28]
\bauthor{\bsnm{Bilar},~\bfnm{D.}\binits{D.}}
(\byear{2007}).
\btitle{Opcodes as predictor for malware}.
\bjournal{International Journal of Electronic Security and Digital Forensics}
\bvolume{1}
\bpages{156--168}.
\end{barticle}
\bptok{imsref}%
\endbibitem

\bibitem[\protect\citeauthoryear{Christodorescu and Jha}{2003}]{Christodorescu03staticanalysis}
\begin{bmisc}[auto:STB|2014/01/06|10:16:28]
\bauthor{\bsnm{Christodorescu},~\bfnm{M.}\binits{M.}} \AND
\bauthor{\bsnm{Jha},~\bfnm{S.}\binits{S.}}
(\byear{2003}).
\bhowpublished{Static analysis of executables to detect malicious patterns.
In \textit{Proceedings of the 12th USENIX Security Symposium}
169--186. USENIX Association, Berkeley, CA.}
\end{bmisc}
\bptok{imsref}%
\endbibitem

\bibitem[\protect\citeauthoryear{Cova, Kruegel and Vigna}{2010}]{Cova10}
\begin{bmisc}[auto:STB|2014/01/06|10:16:28]
\bauthor{\bsnm{Cova},~\bfnm{M.}\binits{M.}},
\bauthor{\bsnm{Kruegel},~\bfnm{C.}\binits{C.}} \AND
\bauthor{\bsnm{Vigna},~\bfnm{G.}\binits{G.}}
(\byear{2010}).
\bhowpublished{Detection and analysis of drive-by-download attacks and
malicious javascript code.
In \textit{Proceedings of the 19th International Conference on World Wide Web}
281--290. ACM, New York.}
\end{bmisc}
\bptok{imsref}%
\endbibitem

\bibitem[\protect\citeauthoryear{Dai, Guha and Lee}{2009}]{Dai09Efficient}
\begin{barticle}[auto:STB|2014/01/06|10:16:28]
\bauthor{\bsnm{Dai},~\bfnm{J.}\binits{J.}},
\bauthor{\bsnm{Guha},~\bfnm{R.}\binits{R.}} \AND
\bauthor{\bsnm{Lee},~\bfnm{J.}\binits{J.}}
(\byear{2009}).
\btitle{Efficient virus detection using dynamic instruction sequences}.
\bjournal{Journal of Computers}
\bvolume{4}
\bpages{405--414}.
\end{barticle}
\bptok{imsref}%
\endbibitem

\bibitem[\protect\citeauthoryear{Dinaburg et~al.}{2008}]{Dinaburg2008}
\begin{bmisc}[auto:STB|2014/01/06|10:16:28]
\bauthor{\bsnm{Dinaburg},~\bfnm{A.}\binits{A.}},
\bauthor{\bsnm{Royal},~\bfnm{P.}\binits{P.}},
\bauthor{\bsnm{Sharif},~\bfnm{M.}\binits{M.}} \AND
\bauthor{\bsnm{Lee},~\bfnm{W.}\binits{W.}}
(\byear{2008}).
\bhowpublished{Ether: Malware analysis via hardware virtualization extensions.
In \textit{Proceedings of the 15th ACM Conference on Computer and Communications Security}
51--62. ACM, New York.}
\end{bmisc}
\bptok{imsref}%
\endbibitem

\bibitem[\protect\citeauthoryear{Efron et~al.}{2004}]{Efron04b}
\begin{barticle}[mr]
\bauthor{\bsnm{Efron},~\bfnm{Bradley}\binits{B.}},
\bauthor{\bsnm{Hastie},~\bfnm{Trevor}\binits{T.}},
\bauthor{\bsnm{Johnstone},~\bfnm{Iain}\binits{I.}} \AND
\bauthor{\bsnm{Tibshirani},~\bfnm{Robert}\binits{R.}}
(\byear{2004}).
\btitle{Least angle regression}.
\bjournal{Ann. Statist.}
\bvolume{32}
\bpages{407--499}.
\bid{doi={10.1214/009053604000000067}, issn={0090-5364}, mr={2060166}}
\bptnote{check related}%
\end{barticle}
\bptok{imsref}%
\endbibitem

\bibitem[\protect\citeauthoryear{Goldberg et~al.}{1996}]{Goldberg96}
\begin{bmisc}[auto:STB|2014/01/06|10:16:28]
\bauthor{\bsnm{Goldberg},~\bfnm{I.}\binits{I.}},
\bauthor{\bsnm{Wagner},~\bfnm{D.}\binits{D.}},
\bauthor{\bsnm{Thomas},~\bfnm{R.}\binits{R.}} \AND
\bauthor{\bsnm{Brewer},~\bfnm{E.}\binits{E.}}
(\byear{1996}).
\bhowpublished{A secure environment for untrusted helper applications (confining the wily hacker).
In \textit{Proceedings of the Sixth USENIX UNIX Security Symposium} \textbf{6}
1. USENIX Association, Berkeley, CA.}
\end{bmisc}
\bptok{imsref}%
\endbibitem

\bibitem[\protect\citeauthoryear{Gramacy and Polson}{2012}]{Gramacy12b}
\begin{barticle}[mr]
\bauthor{\bsnm{Gramacy},~\bfnm{Robert~B.}\binits{R.~B.}} \AND
\bauthor{\bsnm{Polson},~\bfnm{Nicholas~G.}\binits{N.~G.}}
(\byear{2012}).
\btitle{Simulation-based regularized logistic regression}.
\bjournal{Bayesian Anal.}
\bvolume{7}
\bpages{567--589}.
\bid{doi={10.1214/12-BA719}, issn={1936-0975}, mr={2981628}}
\end{barticle}
\bptok{imsref}%
\endbibitem

\bibitem[\protect\citeauthoryear{Hastie and Tibshirani}{1996}]{Hastie96}
\begin{barticle}[mr]
\bauthor{\bsnm{Hastie},~\bfnm{Trevor}\binits{T.}} \AND
\bauthor{\bsnm{Tibshirani},~\bfnm{Robert}\binits{R.}}
(\byear{1996}).
\btitle{Discriminant analysis by {G}aussian mixtures}.
\bjournal{J. R. Stat. Soc. Ser. B Stat. Methodol.}
\bvolume{58}
\bpages{155--176}.
\bid{issn={0035-9246}, mr={1379236}}
\end{barticle}
\bptok{imsref}%
\endbibitem

\bibitem[\protect\citeauthoryear{Hofmeyr, Forrest and Somayaji}{1998}]{Hofmeyr98Intrusion}
\begin{barticle}[auto:STB|2014/01/06|10:16:28]
\bauthor{\bsnm{Hofmeyr},~\bfnm{S.~A.}\binits{S.~A.}},
\bauthor{\bsnm{Forrest},~\bfnm{S.}\binits{S.}} \AND
\bauthor{\bsnm{Somayaji},~\bfnm{A.}\binits{A.}}
(\byear{1998}).
\btitle{Intrusion detection using sequences of system calls}.
\bjournal{Journal of Computer Security}
\bvolume{6}
\bpages{151--180}.
\end{barticle}
\bptok{imsref}%
\endbibitem

\bibitem[\protect\citeauthoryear{King and Zeng}{2001}]{King01}
\begin{barticle}[auto:STB|2014/01/06|10:16:28]
\bauthor{\bsnm{King},~\bfnm{G.}\binits{G.}} \AND
\bauthor{\bsnm{Zeng},~\bfnm{L.}\binits{L.}}
(\byear{2001}).
\btitle{Logistic regression in rare events data}.
\bjournal{Political Analysis}
\bvolume{9}
\bpages{137--163}.
\end{barticle}
\bptok{imsref}%
\endbibitem

\bibitem[\protect\citeauthoryear{Kolter and Maloof}{2006}]{Kolter06Learning}
\begin{barticle}[mr]
\bauthor{\bsnm{Kolter},~\bfnm{J.~Zico}\binits{J.~Z.}} \AND
\bauthor{\bsnm{Maloof},~\bfnm{Marcus~A.}\binits{M.~A.}}
(\byear{2006}).
\btitle{Learning to detect and classify malicious executables in the wild}.
\bjournal{J. Mach. Learn. Res.}
\bvolume{7}
\bpages{2721--2744}.
\bid{issn={1532-4435}, mr={2274458}}
\end{barticle}
\bptok{imsref}%
\endbibitem

\bibitem[\protect\citeauthoryear{Luk et~al.}{2005}]{Luk05}
\begin{bmisc}[auto:STB|2014/01/06|10:16:28]
\bauthor{\bsnm{Luk},~\bfnm{C.-K.}\binits{C.-K.}},
\bauthor{\bsnm{Cohn},~\bfnm{R.}\binits{R.}},
\bauthor{\bsnm{Muth},~\bfnm{R.}\binits{R.}},
\bauthor{\bsnm{Patil},~\bfnm{H.}\binits{H.}},
\bauthor{\bsnm{Klauser},~\bfnm{A.}\binits{A.}},
\bauthor{\bsnm{Lowney},~\bfnm{G.}\binits{G.}},
\bauthor{\bsnm{Wallace},~\bfnm{S.}\binits{S.}},
\bauthor{\bsnm{Reddi},~\bfnm{V.~J.}\binits{V.~J.}} \AND
\bauthor{\bsnm{Hazelwood},~\bfnm{K.}\binits{K.}}
(\byear{2005}).
\bhowpublished{Pin: Building customized program analysis tools
with dynamic instrumentation.
In \textit{Proceedings of the ACM SIGPLAN Conference on
Programming Language Design and Implementation}
190--200. ACM, New York.}
\end{bmisc}
\bptok{imsref}%
\endbibitem

\bibitem[\protect\citeauthoryear{Manski and Lerman}{1977}]{Manski77}
\begin{barticle}[mr]
\bauthor{\bsnm{Manski},~\bfnm{Charles~F.}\binits{C.~F.}} \AND
\bauthor{\bsnm{Lerman},~\bfnm{Steven~R.}\binits{S.~R.}}
(\byear{1977}).
\btitle{The estimation of choice probabilities from choice based samples}.
\bjournal{Econometrica}
\bvolume{45}
\bpages{1977--1988}.
\bid{issn={0012-9682}, mr={0501708}}
\end{barticle}
\bptok{imsref}%
\endbibitem

\bibitem[\protect\citeauthoryear{Meinshausen}{2007}]{Meinshausen07}
\begin{barticle}[mr]
\bauthor{\bsnm{Meinshausen},~\bfnm{Nicolai}\binits{N.}}
(\byear{2007}).
\btitle{Relaxed {L}asso}.
\bjournal{Comput. Statist. Data Anal.}
\bvolume{52}
\bpages{374--393}.
\bid{doi={10.1016/j.csda.2006.12.019}, issn={0167-9473}, mr={2409990}}
\end{barticle}
\bptok{imsref}%
\endbibitem

\bibitem[\protect\citeauthoryear{PandaLabs}{2012}]{Panda12}
\begin{bmisc}[auto:STB|2014/01/06|10:16:28]
\borganization{PandaLabs}
(\byear{2012}).
\bhowpublished{PandaLabs quarterly report. Available at
\texttt{
\href{http://press.pandasecurity.com/wp-content/uploads/2012/08/Quarterly-Report-PandaLabs-April-June-2012.pdf}{http://press.}
\href{http://press.pandasecurity.com/wp-content/uploads/2012/08/Quarterly-Report-PandaLabs-April-June-2012.pdf}{pandasecurity.com/wp-content/uploads/2012/08/Quarterly-Report-PandaLabs-}
\href{http://press.pandasecurity.com/wp-content/uploads/2012/08/Quarterly-Report-PandaLabs-April-June-2012.pdf}{April-June-2012.pdf}}}.
\end{bmisc}
\bptok{imsref}%
\endbibitem

\bibitem[\protect\citeauthoryear{Perdisci et~al.}{2006}]{Perdisci06misleadingworm}
\begin{bmisc}[auto:STB|2014/01/06|10:16:28]
\bauthor{\bsnm{Perdisci},~\bfnm{R.}\binits{R.}},
\bauthor{\bsnm{Dagon},~\bfnm{D.}\binits{D.}},
\bauthor{\bsnm{Fogla},~\bfnm{P.}\binits{P.}} \AND
\bauthor{\bsnm{Sharif},~\bfnm{M.}\binits{M.}}
(\byear{2006}).
\bhowpublished{Misleading worm signature generators using deliberate noise
injection.
In \textit{Proceedings of the IEEE Symposium on Security and Privacy}
17--31. IEEE Computer Society Technical Committee on Security and Privacy.}
\end{bmisc}
\bptok{imsref}%
\endbibitem

\bibitem[\protect\citeauthoryear{Prentice and Pyke}{1979}]{Prentice79}
\begin{barticle}[mr]
\bauthor{\bsnm{Prentice},~\bfnm{R.~L.}\binits{R.~L.}} \AND
\bauthor{\bsnm{Pyke},~\bfnm{R.}\binits{R.}}
(\byear{1979}).
\btitle{Logistic disease incidence models and case--control studies}.
\bjournal{Biometrika}
\bvolume{66}
\bpages{403--411}.
\bid{doi={10.1093/biomet/66.3.403}, issn={0006-3444}, mr={0556730}}
\end{barticle}
\bptok{imsref}%
\endbibitem

\bibitem[\protect\citeauthoryear{Quist}{2012}]{Quistweb}
\begin{bmisc}[auto:STB|2014/01/06|10:16:28]
\bauthor{\bsnm{Quist},~\bfnm{D.}\binits{D.}}
(\byear{2012}).
\bhowpublished{Community malicious code research and analysis. Available at \url{http://www.offensivecomputing.net/}.}
\end{bmisc}
\bptok{imsref}%
\endbibitem

\bibitem[\protect\citeauthoryear{Reddy, Dash and Pujari}{2006}]{Reddy06NewMaliciousCode}
\begin{bincollection}[auto:STB|2014/01/06|10:16:28]
\bauthor{\bsnm{Reddy},~\bfnm{D.~K.~S.}\binits{D.~K.~S.}},
\bauthor{\bsnm{Dash},~\bfnm{S.}\binits{S.}} \AND
\bauthor{\bsnm{Pujari},~\bfnm{A.}\binits{A.}}
(\byear{2006}).
\btitle{New malicious code detection using variable length $n$-grams}.
In \bbooktitle{Information Systems Security}.
\bseries{Lecture Notes in Computer Science}
\bvolume{4332}
\bpages{276--288}.
\bpublisher{Springer},
\blocation{Berlin}.
\end{bincollection}
\bptok{imsref}%
\endbibitem

\bibitem[\protect\citeauthoryear{Reddy and Pujari}{2006}]{Reddy06gramanalysis}
\begin{barticle}[auto:STB|2014/01/06|10:16:28]
\bauthor{\bsnm{Reddy},~\bfnm{D.}\binits{D.}} \AND
\bauthor{\bsnm{Pujari},~\bfnm{A.}\binits{A.}}
(\byear{2006}).
\btitle{$N$-gram analysis for computer virus detection}.
\bjournal{Journal in Computer Virology}
\bvolume{2}
\bpages{231--239}.
\end{barticle}
\bptok{imsref}%
\endbibitem

\bibitem[\protect\citeauthoryear{Rieck et~al.}{2011}]{Rieck11Automatic}
\begin{barticle}[auto:STB|2014/01/06|10:16:28]
\bauthor{\bsnm{Rieck},~\bfnm{K.}\binits{K.}},
\bauthor{\bsnm{Trinius},~\bfnm{P.}\binits{P.}},
\bauthor{\bsnm{Willems},~\bfnm{C.}\binits{C.}} \AND
\bauthor{\bsnm{Holz},~\bfnm{T.}\binits{T.}}
(\byear{2011}).
\btitle{Automatic analysis of malware behavior using machine learning}.
\bjournal{Journal of Computer Security}
\bvolume{19}
\bpages{639--668}.
\end{barticle}
\bptok{imsref}%
\endbibitem

\bibitem[\protect\citeauthoryear{Royal et~al.}{2006}]{Royal06}
\begin{bincollection}[auto:STB|2014/01/06|10:16:28]
\bauthor{\bsnm{Royal},~\bfnm{P.}\binits{P.}},
\bauthor{\bsnm{Halpin},~\bfnm{M.}\binits{M.}},
\bauthor{\bsnm{Dagon},~\bfnm{D.}\binits{D.}},
\bauthor{\bsnm{Edmonds},~\bfnm{R.}\binits{R.}} \AND
\bauthor{\bsnm{Lee},~\bfnm{W.}\binits{W.}}
(\byear{2006}).
\btitle{Polyunpack: Automating the hidden-code extraction of unpackexecuting malware}.
In \bbooktitle{Proceedings of the 22nd Annual Computer Security Applications Conference}
\bpages{289--300}.
\end{bincollection}
\bptok{imsref}%
\endbibitem

\bibitem[\protect\citeauthoryear{Shafiq, Khayam and Farooq}{2008}]{Shafiq08}
\begin{bmisc}[auto:STB|2014/01/06|10:16:28]
\bauthor{\bsnm{Shafiq},~\bfnm{M.}\binits{M.}},
\bauthor{\bsnm{Khayam},~\bfnm{S.}\binits{S.}} \AND
\bauthor{\bsnm{Farooq},~\bfnm{M.}\binits{M.}}
(\byear{2008}).
\bhowpublished{Embedded malware detection using Markov $n$-grams.
In \textit{Proceedings of the 5th International Conference on
Detection of Intrusions and Malware, and Vulnerability Assessment}
88--107. ACM, New York.}
\end{bmisc}
\bptok{imsref}%
\endbibitem

\bibitem[\protect\citeauthoryear{Shankarapani et~al.}{2010}]{Shank2010Malware}
\begin{barticle}[auto:STB|2014/01/06|10:16:28]
\bauthor{\bsnm{Shankarapani},~\bfnm{M.}\binits{M.}},
\bauthor{\bsnm{Ramamoorthy},~\bfnm{S.}\binits{S.}},
\bauthor{\bsnm{Movva},~\bfnm{R.}\binits{R.}} \AND
\bauthor{\bsnm{Mukkamala},~\bfnm{S.}\binits{S.}}
(\byear{2010}).
\btitle{Malware detection using assembly and API call sequences}.
\bjournal{Journal in Computer Virology}
\bvolume{7}
\bpages{1--13}.
\end{barticle}
\bptok{imsref}%
\endbibitem

\bibitem[\protect\citeauthoryear{Skaletsky et~al.}{2010}]{Skaletsky10}
\begin{bmisc}[auto:STB|2014/01/06|10:16:28]
\bauthor{\bsnm{Skaletsky},~\bfnm{A.}\binits{A.}},
\bauthor{\bsnm{Devor},~\bfnm{T.}\binits{T.}},
\bauthor{\bsnm{Chachmon},~\bfnm{N.}\binits{N.}},
\bauthor{\bsnm{Cohn},~\bfnm{R.}\binits{R.}},
\bauthor{\bsnm{Hazelwood},~\bfnm{K.}\binits{K.}},
\bauthor{\bsnm{Vladimirov},~\bfnm{V.}\binits{V.}} \AND
\bauthor{\bsnm{Bach},~\bfnm{M.}\binits{M.}}
(\byear{2010}).
\bhowpublished{Dynamic program analysis of Microsoft Windows
applications. In \textit{2010 International Symposium on Performance
Analysis of Software and Systems} (\textit{ISPASS}) 2--12.
IEEE Computer Society's Technical Committee on the Internet}.
\end{bmisc}
\bptok{imsref}%
\endbibitem

\bibitem[\protect\citeauthoryear{Stolfo, Wang and Li}{2007}]{Stolfo07TowardsStealthy}
\begin{bincollection}[auto:STB|2014/01/06|10:16:28]
\bauthor{\bsnm{Stolfo},~\bfnm{S.}\binits{S.}},
\bauthor{\bsnm{Wang},~\bfnm{K.}\binits{K.}} \AND
\bauthor{\bsnm{Li},~\bfnm{W.-J.}\binits{W.-J.}}
(\byear{2007}).
\btitle{Towards stealthy malware detection}.
In \bbooktitle{Malware Detection}.
\bseries{Advances in Information Security}
\bvolume{27}
\bpages{231--249}.
\bpublisher{Springer},
\blocation{New York}.
\end{bincollection}
\bptok{imsref}%
\endbibitem

\bibitem[\protect\citeauthoryear{Storlie et~al.}{2014}]{Storlie13malwaresupp}
\begin{bmisc}[auto:STB|2014/01/06|10:16:28]
\bauthor{\bsnm{Storlie},~\bfnm{C.}\binits{C.}},
\bauthor{\bsnm{Anderson},~\bfnm{B.}\binits{B.}},
\bauthor{\bsnm{Vander Wiel},~\bfnm{S.}\binits{S.}},
\bauthor{\bsnm{Quist},~\bfnm{D.}\binits{D.}},
\bauthor{\bsnm{Hash},~\bfnm{C.}\binits{C.}} \AND
\bauthor{\bsnm{Brown},~\bfnm{N.}\binits{N.}}
(\byear{2014}).
\bhowpublished{Supplement to
``Stochastic identification of malware with dynamic~traces.''
DOI:\doiurl{10.1214/13-AOAS703SUPP}.}
\end{bmisc}
\bptok{imsref}%
\endbibitem

\bibitem[\protect\citeauthoryear{Symantec}{2008}]{Symantec2008}
\begin{bmisc}[auto:STB|2014/01/06|10:16:28]
\borganization{Symantec}
(\byear{2008}).
\bhowpublished{Internet security threat report, trends for July--December 2007 (executive summary). White paper. Available at
\texttt{\href{http://eval.symantec.com/mktginfo/enterprise/white\_papers/b-whitepaper\_exec\_summary\_internet\_security\_threat\_report\_xiii\_04-2008.en-us.pdf}{http://eval.symantec.com/mktginfo/}
\href{http://eval.symantec.com/mktginfo/enterprise/white\_papers/b-whitepaper\_exec\_summary\_internet\_security\_threat\_report\_xiii\_04-2008.en-us.pdf}{enterprise/white\_papers/b-whitepaper\_exec\_summary\_internet\_security\_threat\_}
\href{http://eval.symantec.com/mktginfo/enterprise/white\_papers/b-whitepaper\_exec\_summary\_internet\_security\_threat\_report\_xiii\_04-2008.en-us.pdf}{report\_xiii\_04-2008.en-us.pdf}}
}.
\end{bmisc}
\bptok{imsref}%
\endbibitem

\bibitem[\protect\citeauthoryear{Symantec}{2011}]{Symantec2010}
\begin{bmisc}[auto:STB|2014/01/06|10:16:28]
\borganization{Symantec}
(\byear{2011}).
\bhowpublished{Internet security threat report, volume 16. White paper. Available at
\url{http://www.symantec.com/business/threatreport/index.jsp}}.
\end{bmisc}
\bptok{imsref}%
\endbibitem

\bibitem[\protect\citeauthoryear{Taddy}{2013}]{Taddy13}
\begin{barticle}[auto:STB|2014/01/06|10:16:28]
\bauthor{\bsnm{Taddy},~\bfnm{M.}\binits{M.}}
(\byear{2013}).
\btitle{Multinomial inverse regression for text analysis}.
\bjournal{J. Amer. Statist. Assoc.}
\bvolume{108}
\bpages{755--770}.
\end{barticle}
\bptok{imsref}%
\endbibitem

\bibitem[\protect\citeauthoryear{Tibshirani}{1996}]{Tibs96}
\begin{barticle}[mr]
\bauthor{\bsnm{Tibshirani},~\bfnm{Robert}\binits{R.}}
(\byear{1996}).
\btitle{Regression shrinkage and selection via the Lasso}.
\bjournal{J. R. Stat. Soc. Ser. B Stat. Methodol.}
\bvolume{58}
\bpages{267--288}.
\bid{issn={0035-9246}, mr={1379242}}
\end{barticle}
\bptok{imsref}%
\endbibitem

\bibitem[\protect\citeauthoryear{Zou}{2006}]{Zou06b}
\begin{barticle}[mr]
\bauthor{\bsnm{Zou},~\bfnm{Hui}\binits{H.}}
(\byear{2006}).
\btitle{The Adaptive Lasso and its oracle properties}.
\bjournal{J. Amer. Statist. Assoc.}
\bvolume{101}
\bpages{1418--1429}.
\bid{doi={10.1198/016214506000000735}, issn={0162-1459}, mr={2279469}}
\end{barticle}
\bptok{imsref}%
\endbibitem

\bibitem[\protect\citeauthoryear{Zou and Hastie}{2005}]{Zou05}
\begin{barticle}[mr]
\bauthor{\bsnm{Zou},~\bfnm{Hui}\binits{H.}} \AND
\bauthor{\bsnm{Hastie},~\bfnm{Trevor}\binits{T.}}
(\byear{2005}).
\btitle{Regularization and variable selection via the elastic net}.
\bjournal{J. R. Stat. Soc. Ser. B Stat. Methodol.}
\bvolume{67}
\bpages{301--320}.
\bid{doi={10.1111/j.1467-9868.2005.00503.x}, issn={1369-7412}, mr={2137327}}
\end{barticle}
\bptok{imsref}%
\endbibitem

\end{thebibliography}
\end{document}